# Field-free spin-orbit switching of canted magnetization in Pt/Co/Ru/RuO$_2$(101) multilayers

Yunzhuo Wu[1], Tong Wu[1], Haoran Chen[1], Yongwei Cui[1], Hongyue Xu[1], Nan Jiang[1], Zhen Cheng[1], Yizheng Wu[1,2,3, *]

[1]Department of Physics and State Key Laboratory of Surface Physics, Fudan University, Shanghai 200433, China

[2]Shanghai Research Center for Quantum Sciences, Shanghai 201315, China

[3]Shanghai Key Laboratory of Metasurfaces for Light Manipulation, Fudan University, Shanghai 200433, China

## Abstract

Enabling field-free current-induced switching of perpendicular magnetization is essential for advancing spin-orbit-torque magnetic random access memory technology. Our research on the Pt/Co/Ru/RuO$_2$(101) system has successfully demonstrated field-free switching through current injection along the RuO$_2$[010] axis. We discovered that the system exhibits a tilted easy axis, inclined from the out-of-plane towards the RuO$_2$[$\bar{1}$01] direction. The application of current perpendicular to this tilted axis generates a substantial out-of-plane effective field, which facilitates field-free magnetization switching. Our results also indicate that adjusting the thickness of the Ru layer to optimize the tilt angle can significantly reduce the critical switching current density. This work provides a viable strategy for controlling the tilting magnetization, essential for the development of RuO$_2$-based magnetic devices.

Spin–orbit torque (SOT) is a crucial technology for the next-generation magnetic random access memory (MRAM), facilitating efficient manipulation of magnetization in spintronic devices[1,2]. The pursuit of low writing current density, minimal power consumption, and nonvolatility through full electrical control is central to SOT-MRAM research. The manipulation of magnetization via SOT typically depends on spin currents generated by substantial spin-orbit coupling (SOC) in heavy metals (HMs). Achieving SOT switching of magnetization with perpendicular magnetic anisotropy (PMA) has typically required an external magnetic field aligned with the current direction to break the symmetry[3,4]. This necessity for an additional in-plane field poses a challenge for practical device applications. Various innovative methods have been explored to facilitate field-free magnetization switching[5-8]. These include breaking the structural and magnetic symmetry in samples with wedge-shaped structure[9,10], utilizing a gradient oxidized capping layer[8], and employing exchange bias[11-14]. Moreover, the out-of-plane SOT can be induced in materials with broken symmetry, such as $WTe_2$[15], $Mn_2Au$[16] and CuPt/CoPt bilayers[17,18], when the current is injected along the low-symmetry axis of the nonmagnetic layers. These strategies are crucial steps towards the development of more efficient and versatile SOT-MRAM devices.

$RuO_2$, with its rutile crystal structure, is an emerging antiferromagnetic material that has garnered attention in the field of spintronics. The spin-splitting effect (SSE) in $RuO_2$, which results from its spin-splitting band, has been theoretically predicted and remains a subject of ongoing research[19-21]. When a charge current is injected, the spin polarization induced by the SSE is found to align with the Néel vector direction of $RuO_2$[22-24]. The anisotropic SSE has been experimentally observed using the spin-torque ferromagnetic resonance method, which confirmed the presence of a spin current with polarization along the z-direction ($\sigma_z$)[22,23]. S. Karube et al. have reported the field-free magnetization switching behavior in the Pt/Co/Ru/$RuO_2$(101) multilayer stacks[22]. Additionally, in the Ru/Pt/Co/Pt/$RuO_2$(101) heterostructures, robust field-free switching can be realized induced by spin current with $\sigma_z$[25]. However, recent muon spin rotation and relaxation experiments suggested that bulk $RuO_2$ behaves as a non-magnetic metal[26]. Additionally, Z.Q. Wang et al. reported that the anisotropic inverse spin Hall effect is the dominant mechanism for spin-charge conversion in the $RuO_2$(101) films[27]. These findings raise questions about whether the field-free switching behavior observed in the Pt/Co/Ru/$RuO_2$(101) system is indeed induced by the SSE. Therefore, further intensive research is necessary to investigate the origin of the field-free SOT-induced magnetization switching in $RuO_2$-based magnetic systems.

In this letter, we systematically investigated the field-free magnetization switching behavior in $RuO_2$-based magnetic system. We fabricated a Pt/Co/Ru multilayer exhibiting PMA on the $RuO_2$(101) plane. Consistent field-free switching can be achieved for the current injected along the $RuO_2$[010] direction. Our detailed analysis revealed that the magnetic easy axis is not aligned perpendicularly to the plane but is instead tilted, with its in-plane component oriented towards $RuO_2$[$\bar{1}$01] direction. This observation suggests that the field-free switching in this structure is predominantly due to the presence of this tilted easy axis. Additionally, measurements under direct current (DC) conditions indicate that the out-of-plane effective field is significant for the current injecting along the [010] direction, while it is almost negligible for the current injecting along the [$\bar{1}$01] direction. Furthermore, it was discovered that optimizing the tilt angle by adjusting the thickness of the Ru layer can significantly reduce the critical switching current density. This reduction is approximately one order of magnitude lower than that typically observed in Co/Pt systems[3,12,28]. These results provide valuable insights into the mechanisms underlying field-free magnetization switching in $RuO_2$-based systems and have significant implications for the development of advanced magnetic devices.

The Pt/Co/Ru/$RuO_2$ films were grown on $Al_2O_3$(1$\bar{1}$02) substrates in a magnetron sputtering system with a base pressure of $2\times10^{-8}$ Torr[29]. The sapphire substrates were pre-annealed at 500°C for 1 hour in the vacuum chamber to prepare the surface for deposition. The $RuO_2$ films were epitaxially deposited through reactive sputtering in an oxygen-argon atmosphere of 3 mTorr, with a ratio of 1:4. Subsequent layers were deposited under an argon pressure of 3 mTorr at room temperature.

X-ray diffraction measurements confirmed the formation of a single crystalline $RuO_2$(101) film[22,29]. To systematically investigate the impact of the intermediate Ru layer on SOT switching, a wedge-shaped Ru layer was introduced, facilitating a gradual change in Ru thickness. The thickness gradient, spanning from 0 to 1.2 nm at a rate of ~ 0.2 nm/mm, was controlled by moving a shutter over the substrate during deposition. Additionally, a thick Ru layer with a thickness of 6.2 nm was also grown as a shoulder on the same sample.

For the electrical measurements, the samples were patterned into Hall bar devices using standard photolithography and Ar+ ion milling techniques. Each device has dimensions of 15 μm in width and 100 μm in length. During the current-induced magnetization switching measurements, a 100 μs current pulse was applied into the Hall bar devices. After each pulse, the Hall resistance was measured 1 second later using a small DC current of 1 mA. A commercial magneto-optic Kerr microscope from Evico Magnetics GmbH was used to investigate the evolution of the magnetic domain images excited by the current pulses[30,31]. To further ascertain the tilt angle and anisotropy field, anomalous Hall effect (AHE) measurements were conducted by varying the magnetic field direction in a superconducting vector magnet system.

The current-induced magnetization switching measurements were initially conducted on a Pt(2 nm)/Co(0.8 nm)/Ru(0.8 nm)/RuO$_2$(8 nm) multilayer stack, which mirrors the sample structure reported in Ref. 22. Fig. 1(a) presents the simultaneous measurement setup for Hall voltage signals and magnetic domain images captured through Kerr microscopy. The presence of PMA in the sample was verified by the characteristic out-of-plane hysteresis loop shown in Fig. 1(b).

Subsequently, the behavior of current-driven magnetization switching under zero external field was investigated. Fig. 1(c) illustrates a clear field-free switching behavior when the current was applied along the [010] direction (*I*//RuO$_2$[010]), with no switching observed for *I*//RuO$_2$[$\bar{1}$01]. Fig. 1(d) displays the domain images post-application of 32 mA current pulses along the [010] direction, which clearly depict full switching at zero field. Thus, the deterministic field-free switching in the Pt/Co/Ru/RuO$_2$(101) system can be conclusively demonstrated, consistent with the results reported in Ref. 22[22].

Fig. 1(e) displays the typical current-induced magnetization switching loops under varying in-plane field strengths for the current directed along the [010] direction. The switching polarity consistently follows a counterclockwise direction, with no changes in polarity for the field up to -1000 Oe. In most systems exhibiting field-free magnetization switching, the switching polarity can be altered with the in-plane field strength exceeding a certain threshold, typically no more than approximately 500 Oe[9,10,32-34]. Thus, our measurements indicate that the magnetization switching in the Pt/Co/Ru/RuO$_2$ system is notably resilient to external field disturbances.

Usually, field-free switching in RuO$_2$-based magnetic systems is attributed to the spin-splitter effect[22]. However, in the Py/RuO$_2$(101) system with PMA, field-free magnetization switching is not observed despite the expected presence of the spin-splitter effect[35]. It is noteworthy that the hysteresis loop depicted in Fig. 1(b) shows no saturation in the AHE signal after magnetization switching, suggesting that the easy axis is tilted away from the z-axis. It has been reported that field-free switching is feasible in systems exhibiting tilted anisotropy when a current is applied perpendicular to the tilted axis[9,36,37]. Therefore, the orientation of the magnetic easy axis and its influence on field-free switching were further investigated.

We subsequently conducted measurements of the hysteresis loop over a broader field range. Fig. 2(a) clearly shows that the Hall signal ($R_H$) saturates at fields exceeding 15 kOe, and the inset $R_H$ loop under the narrow field range exhibits a shape similar to that in Fig. 1(b). This finding unequivocally indicates that the magnetization at zero field is not aligned perpendicularly to the film plane but is inclined from the normal direction. By determining the ratio of the remanent value ($R_{Rem}$) at zero field to the saturation value ($R_{Sat}$), it was deduced that the easy axis is inclined at an angle of approximately 66º from the z-axis.

The $R_H$ measurements are only sensitive to the z-component of magnetization, which is insufficient to ascertain the in-plane orientation of the easy axis. Thus, we proceeded to measure the hysteresis loops with fields applied in different directions utilizing a superconducting quantum interference device (SQUID). Fig. 2(b) displays the in-plane and out-of-plane magnetization hysteresis (M-H) loops. The inset indicates the presence of remanent magnetization in both the [$\bar{1}$01] direction and the z-direction, yet the absence in the [010] direction. The remanent magnetization along the [$\bar{1}$01] direction surpasses that along the z-direction, implying that the easy axis of the magnetic moment deviates from the z-axis towards the [$\bar{1}$01] direction, as depicted in Fig. 2(c).

Both experimental evidence and theoretical models have confirmed that field-free switching is achievable in systems with a tilted easy axis[7,9,36-38]. This type of switching is facilitated for the current injected perpendicular to the tilted axis. In the system under investigation, the in-plane component of the tilted magnetization is oriented along the [$\bar{1}$01] direction, and field-free switching is observed for $I$//RuO$_2$[010] (Fig. 1(c)). Thus, we conclude that the field-free switching in this system is primarily driven by the tilting of the easy axis, suggesting that the SSE may not significantly contribute to the observed field-free switching behavior. It is noted that the ultrathin Py film grown on RuO$_2$(101) surface also exhibits strong PMA, yet no field-free switching is detected[35], which further supports the notion that the SSE in Pt/Co/Ru/RuO$_2$(101) system does not play a crucial role in the current-induced magnetization switching.

In the magnetic system with a tilted easy axis, it is well-documented that an out-of-plane effective field can be induced by the current injected perpendicular to the tilted axis[7,9,36,37]. However, when current is injected parallel to the tilted axis, no significant out-of-plane effective field can be observed. To evaluate the out-of-plane effective field induced by SOT on the tilted magnetization, we performed the hysteresis loop measurements with varying DC current strengths [13,17,18].

Figs. 3(a) and (b) depict the typical hysteresis loops under different currents for $I$//RuO$_2$[010] and $I$//RuO$_2$[$\bar{1}$01], respectively. At currents below 15 mA, the hysteresis loops for both current polarities are identical, suggesting the absence of a notable out-of-plane effective field. At a current of 17 mA with $I$//RuO$_2$[010], the hysteresis loop shifts completely to the left for positive current and to the right for negative current. This shift indicates the presence of an out-of-plane effective field, which serves as the driving force for the observed deterministic field-free switching. It is noted that the current-induced switching in Fig. 1(c) requires a significantly higher current strength above 24 mA for pulsed current, potentially due to disparate heating effects under DC and pulsed current conditions. In contrast, for the case with 17 mA current applied along the [$\bar{1}$01] direction, as shown in Fig. 3(b), the hysteresis loops for both positive and negative currents are similar, indicating a minimal out-of-plane effective field. This is in line with the observation that field-free switching does not occur for $I$//RuO$_2$[$\bar{1}$01].

Fig. 3(c) presents the current-dependent coercivities ($H_c$) of the hysteresis loops for both current directions. The $H_c$ exhibit a similar decline with current for $I$<15 mA, which is attributable to the effect of Joule heating. However, beyond 15 mA, the $H_c$ for $I$//RuO$_2$[010] is markedly lower than that for $I$//RuO$_2$[$\bar{1}$01]. This disparity is due to the influence of the current-induced effective field. Fig. 3(d) illustrates the shift field $H_{\text{shift}}$ as a function of current. For $I$//RuO$_2$[010], $H_{\text{shift}}$ exhibits opposite signs for opposite current directions when the current exceeds 15 mA. The maximum measurable $H_{\text{shift}}$ is observed to reach 100 Oe at a current of 17 mA. In contrast, for $I$//RuO$_2$[$\bar{1}$01], $H_{\text{shift}}$ remains negligible across all applied current levels.

We have further quantified the out-of-plane effective field $H_z^{\text{eff}}$ as a function of the averaged current density, as depicted in the inset of Fig. 3(d). $H_z^{\text{eff}}$ is determined by averaging the $H_{\text{shift}}$ under positive and negative currents, i.e. $H_z^{\text{eff}} = |H_{\text{shift}}(+I) - H_{\text{shift}}(-I)|/2$. The current density is derived by dividing the current by the total cross-sectional area of the multilayer stack. At a current of 17 mA, the average current density is $9.7 \times 10^{10}$ A/m$^2$, which generates $H_z^{\text{eff}}$ of 100 Oe. Consequently, the out-of-plane SOT efficiency is calculated to be $\chi = H_z^{\text{eff}}/J = 103$ Oe/$10^{11}$Am$^{-2}$. This efficiency significantly exceeds the previously determined values of 3 Oe/$10^{11}$Am$^{-2}$ in the Pt/Co/Pt system[9], and 56 Oe/$10^{11}$Am$^{-2}$ in the TaO$_x$/CoFeB/Ta system[8].

It is important to note that current densities in the RuO$_2$ layer and other metal layers may vary due to different resistivities across layers. In Ref. 22, the determined resistivity is 160 μΩ cm in the Pt layer and 55 μΩ cm in the RuO$_2$ layer. The higher resistivity and the thinner thickness of the Pt layer implies a relatively higher current density in RuO$_2$, suggesting that the RuO$_2$ layer is the primary contributor to the spin current. Assuming that the resistivities of the Ru and Co ultrathin layers are similar to that of the Pt layer, the estimated out-of-plane SOT efficiency due to the current in the RuO$_2$ layer is approximately to be 65 Oe/$10^{11}$Am$^{-2}$, which is still larger than the reported values in other systems[8,9,17]. Furthermore, considering that the Co/Pt system exhibits the same polarity of current-induced switching as observed in the current Pt/Co/Ru/RuO$_2$ system[37], the top Pt layer is expected to have an SOT effect opposite to that of the bottom RuO$_2$ layer. In the Pt/Co/Pt/RuO$_2$ structure in Ref. 25, the upper and lower layers of Pt are of equal thickness, suggesting that the spin current generated by Pt cancels out[25]. The switching behavior induced by the RuO$_2$ is also consistent with our findings. So, the RuO$_2$ layer could possess a very strong SOT effect, highlighting its critical role in the field-free switching performance in this system.

Next, we focus on the role of the Ru interlayer on the field-free switching. Our experiments revealed that that a Pt/Co bilayer directly grown on the RuO$_2$ surface exhibits only in-plane anisotropy. This observation suggests that the Ru interlayer is instrumental in enhancing the PMA in our system. Moreover, the crystal structure of RuO$_2$(101) film shows the mirror-symmetry broken along the [$\bar{1}$01] direction, thus inducing the tilted magnetic anisotropy. It is hypothesized that the tilted magnetic anisotropy varies with the Ru thickness $d_{Ru}$. To delve deeper into the relationship between field-free switching and the angle of the tilted easy axis, we systematically varied $d_{Ru}$ by growing a wedge-shaped Ru layer. This approach enabled us to continuously modulate the tilt angle of the Co magnetic moment, as shown in Fig. 4(a). A series of Hall bar devices with varying $d_{Ru}$ were fabricated on the wedge-shaped sample. To ascertain the tilt angle Θ of the easy axis and the magnetic anisotropy field for different $d_{Ru}$, we employed the AHE measurements with a rotating field in Hall bar devices. This technique is grounded in the principle of magnetic torquemetry, and provides a precise method of measuring the magnetic anisotropy and the orientation of the easy axis in our samples[39,40].

Fig. 4(b) displays the $\theta_H$-dependent R$_H$ curves for different $d_{Ru}$ by rotating the magnetic field, which is maintained at a fixed strength of 15 $k$Oe. The field rotation is conducted in the plane containing the z-axis and the [$\bar{1}$01] direction, with $\theta_H$ representing the field angle, as depicted in the inset of Fig. 4(b). In the $R_H(\theta_H)$ curves for systems with $d_{Ru}$ = 0.53 nm and 0.98 nm, the angles corresponding to the maximum $R_H$ signals clearly deviate from 0 and 180°. This deviation indicates that the easy axis is tilted away from the z-axis. The measured $R_H(\theta_H)$ curve for $d_{Ru}$ = 0 nm exhibits symmetry around 180°, affirming that its easy axis resides in the film plane, as confirmed by subsequent fitting analyses.

The $R_H$ signal is known to be proportional to the z-component of magnetization, i. e. $R_H \propto \cos\theta_m$, which allows us to establish the relationship between the magnetization angle $\theta_m$ and the field angle $\theta_H$[35,39-41]. Then, the anisotropy torque field $\tau(\theta_m)$ can be determined as a function of $\theta_m$ by the formula $\tau(\theta_m) = H sin(\theta_H - \theta_m) = H_k \sin(2(\theta_m - \Theta))$ [39-41]. The uniaxial anisotropy field $H_k$ and the tilt angle $\Theta$ of the magnetic anisotropy can be extracted through fitting the $\tau(\theta_m)$ curve (not shown here). As depicted in Fig. 4(c), the tilt angle gradually changes from 90º to 69º as $d_{Ru}$ increases from 0 nm to 1 nm, and then it gradually decreases for thicker Ru layers, reaching 65° at $d_{Ru}$=6.2 nm. Fig. 4(d) shows that $H_k$ increases steadily with $d_{Ru}$, thereby demonstrating a continuous decrease in $H_k$ with the increasing tilt angle $\Theta$, as depicted in the inset of Fig. 4(d).

We further performed the out-of-plane $R_H$-loop measurements for the Hall devices with varying Ru thicknesses. Fig. 4(e) shows that all loops for $d_{Ru} > 0$ nm exhibit noticeable jumping behavior under a small magnetic field, and the coercivity $H_c$ gradually increases with $d_{Ru}$, as shown in Fig. 4(f). The remanent signals in $R_H$ loops show clear increase with the film thickness, and the saturation signals in Fig. 4(b) have the similar value, thus the hysteresis loop measurements also indicate the z-component of magnetization increase with $d_{Ru}$.

To elucidate the impact of the tilted anisotropy on field-free switching, we performed the current-induced magnetization switching measurements at zero field with $I$//RuO$_2$[010]. As shown in Fig. 4(g), all Hall devices with different $d_{Ru}$ in the wedge-shaped sample, excluding those at $d_{Ru}$=0 nm and $d_{Ru}$=6.2 nm, demonstrate clear field-free switching behavior. By examining the current-induced resistance changes in conjunction with the remanence $R_H$ signals in hysteresis loop presented in Fig. 4(e), we can ascertain that complete magnetization switching is attainable for all the samples illustrated in Fig. 4(g). This comparison highlights the pivotal role of the tilted anisotropy in enabling efficient field-free switching in these devices.

The current-induced switching measurements depicted in Fig. 4(g) also reveal a significant variation in the critical switching current, which is notably low at 3.5 mA for the device with $d_{Ru}$=0.18 nm. Fig. 4(h) presents the critical switching current density $J_{sw}$ as a function of $d_{Ru}$, exhibiting an order of magnitude change as $d_{Ru}$ increases from 0.18 nm to 1.2 nm. The minimum $J_{sw}$ is only $2.2\times10^{10}$ A/m$^2$ observed in the device with $d_{Ru}$=0.18 nm, which has the tilt angle $\Theta$ of 85º. Even when accounting for the shunting effect due to different resistivities across layers, the critical switching current density in this system remains an order of magnitude lower than that typically found in most Pt/Co systems[3,12,28].

This reduced critical switching current density can be ascribed to several factors. Firstly, the anisotropy field strength is considerably lower in devices with a thinner Ru layer, as evidenced in Fig. 4(d). Secondly, the presence of a thin Ru layer between Co and RuO$_2$ may diminish the spin current originating from the RuO$_2$ layer due to the short diffusion length within the Ru layer[42], suggesting that the spin current may have higher transmission in devices with a thinner Ru layer. It is noteworthy that field-free switching is not observed in the device with a 6.2 nm Ru interlayer, despite the magnetic anisotropy being closely comparable to that of devices with a 1.2 nm Ru interlayer. This finding further supports that a strong spin current is generated from the RuO$_2$ layer but can be obstructed by a thick Ru layer.

Moreover, it should be noted that the conventional spin current with polarization along the y-direction ($\sigma_y$) continues to significantly influence current-induced magnetization switching, particularly in devices with current applied along the RuO$_2$[010] direction with in-plane fields. To date, there is a relative scarcity of research on the interplay between $\sigma_y$ and $\sigma_z$ on the current-induced switching. Our study suggests that controlling the magnetization tilting could offer a valuable strategy to optimize current-induced switching, which is crucial for the development of next-generation SOT-MRAM devices.

In summary, we demonstrated that the Pt/Co/Ru multilayers grown on RuO$_2$(101) can realize the field-free current induced magnetization switching for the current injecting in RuO$_2$[010] direction. Our magnetic measurements revealed that the Co layer exhibits a magnetic anisotropy easy axis tilted towards the [$\bar{1}$01] direction. The observed field-free switching behavior is primarily attributed to this tilted magnetic easy axis, which can generate a significant out-of-plane effective field when current is applied along the $RuO_2$[010] direction. In addition, by adjusting the thickness of the inserted Ru layer, we can control the tilt angle of the easy axis. We achieved the field-free switching at a low current density of $2.2\times10^{10}$A/m$^2$ at a tilt angle of 85. The ability to control the tilt angle of the easy axis and achieve low current density switching represents a significant step towards the practical application of these systems in next-generation spintronic devices. Our findings not only offer insights into the mechanisms of field-free switching in $RuO_2$-based magnetic systems but also pave the way for developing more efficient and reliable magnetic memory devices.

The work was supported by the National Key Research and Development Program of China (2022YFA1403300), the National Natural Science Foundation of China (Grant No. 11974079, No. 12274083, and No. 12174028), the Shanghai Municipal Science and Technology Major Project (Grant No. 2019SHZDZX01), and the Shanghai Municipal Science and Technology Basic Research Project (Grant No. 22JC1400200 and No. 23dz2260100).

**Data Availability Statement**

The data that support the findings of this study are available from the corresponding author upon reasonable request.

* wuyizheng@fudan.edu.cn

# Figure 1

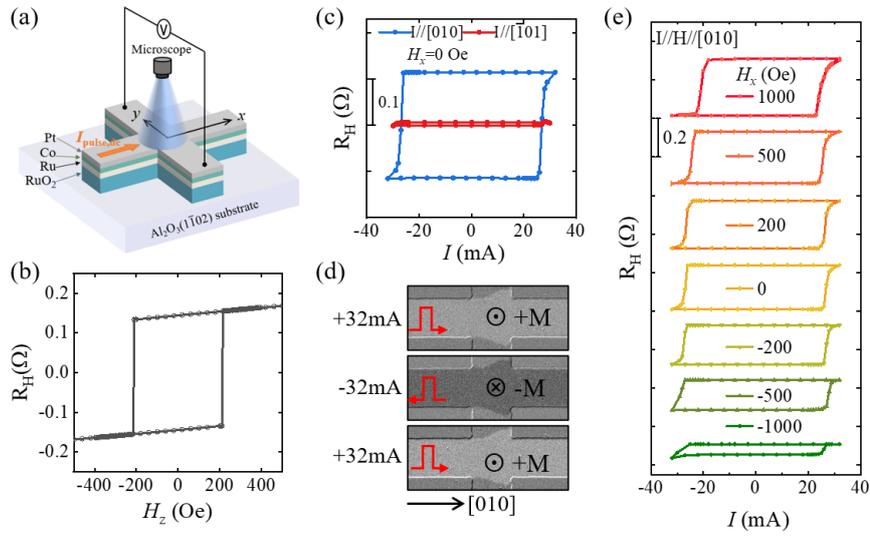

Fig. 1. (a) Experimental geometry for the current-induced switching measurement utilizing a Kerr microscope. (b) Representative AHE hysteresis loop for the Pt(2 nm)/Co(0.8 nm)/Ru(0.8 nm)/RuO$_2$(8 nm)/Al$_2$O$_3$(sub.) sample. (c) Current-induced magnetization switching loops measured at zero field with current applied along the [010] and [$\bar{1}$01] directions. (d) Typical domain images captured after injecting current pulses in the [010] direction, demonstrating the full deterministic current-induced magnetization switching. (e) Current-induced magnetization switching loops assisted by in-plane magnetic fields of various amplitudes and directions.

**Figure 2**

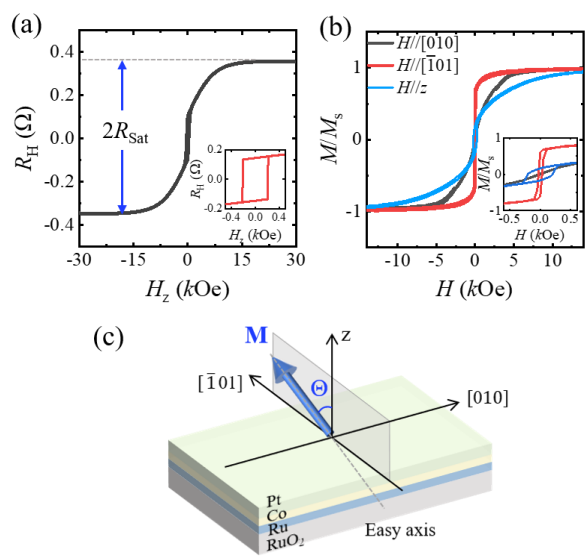

Fig. 2. (a) The typical AHE hysteresis loop of the Pt/Co/Ru/RuO$_2$ multilayer stack measured over a broad magnetic field range. The inset shows the loop measured over a narrow field range. (b) Out-of-plane ($H//z$ axis) and in-plane ($H//[010]$ and $H//[\bar{1}01]$) magnetic hysteresis loops for the Pt/Co/Ru/RuO$_2$ film. The inset indicates the loops in a narrow field range. (c) Schematic representation of the magnetic easy axis orientation in the Co layer, illustrating that it is tilted away from the z-axis towards the RuO$_2$[$\bar{1}$01] direction.

**Figure 3**

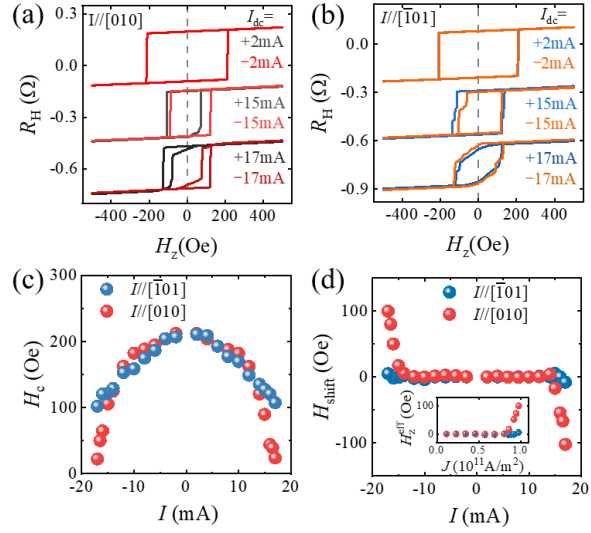

Fig. 3. (a) and (b) Hysteresis loops under ±$I$ for (a) $I//RuO_2[010]$ and (b) $I//RuO_2[\bar{1}01]$, respectively. (c) Variation of the coercive field with the DC currents applied along two directions. (d) Current dependence of the out-of-plane effective field for currents in the [010] (black dot) and [$\bar{1}$01] (red dot) directions. $H_{shift}$ is estimated from the loop shift in (a) and (b). The inset shows the out-of-plane effective field $H_z^{eff}$ as a function of current density.

**Figure 4**

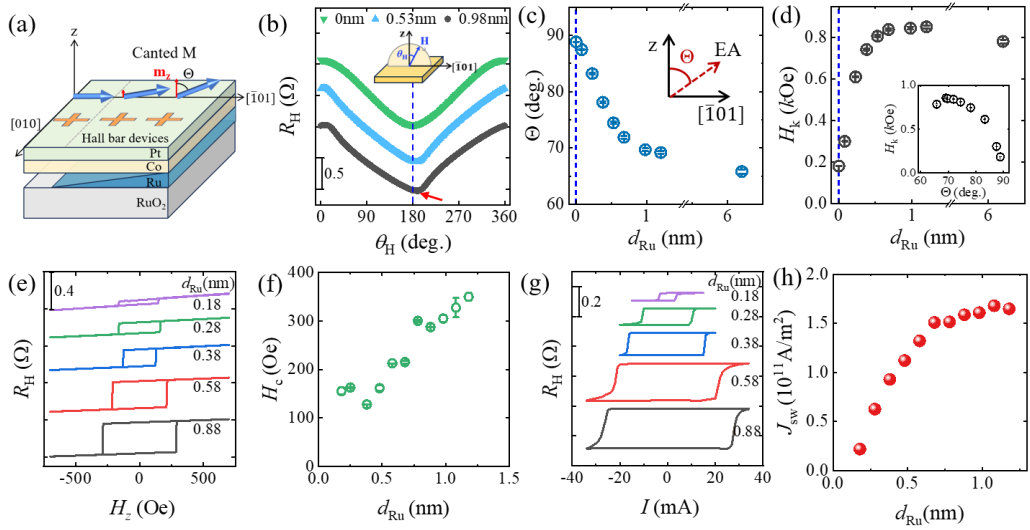

Fig. 4. (a) Schematic drawing of the Pt(2nm)/Co(0.8nm)/Ru($d_{Ru}$)/RuO$_2$(8 nm)/Al$_2$O$_3$(sub.) sample structure with the Ru layer grown into the wedged shape. The canting angle of the magnetic anisotropy also gradually changes with $d_{Ru}$. (b) Typical $\theta_H$-dependent AHE signals for different $d_{Ru}$. The red arrow highlights that the maximum AHE signal occurs at the field angle away from z-axis. (c) Relationship between $d_{Ru}$ and the corresponding tilt angle $\Theta$. (d) Dependence of anisotropy field $H_k$ on $d_{Ru}$. The inset shows $H_k$ as a function of $\Theta$. (e) Typical AHE loops for devices with varying $d_{Ru}$. (f) $H_c$ as a function of $d_{Ru}$, extracted from the AHE loops in (e). (g) Current-induced magnetization switching loops at the zero-field condition for the devices with different $d_{Ru}$. (h) Critical switching current density $J_{sw}$ as a function of $d_{Ru}$.